\documentclass[aps,prb,showpacs,twocolumn]{revtex4-1}
\usepackage{amssymb}
\usepackage{amsmath}
\usepackage{graphicx}
\usepackage{dsfont}
\usepackage{times}

\begin{document}

\title{Critical Level Statistics at the Many-Body Localization Transition
Region}
\author{Wen-Jia Rao}
\email{wjrao@hdu.edu.cn}
\affiliation{School of Science, Hangzhou Dianzi University, Hangzhou 310027, China.}
\date{\today }

\begin{abstract}
We study the critical level statistics at the many-body localization (MBL)
transition region in random spin systems. By employing the inter-sample
randomness as indicator, we manage to locate the MBL transition point in
both orthogonal and unitary models. We further count the $n$-th order gap
ratio distributions at the transition region up to $n=4$, and find they fit
well with the short-range plasma model (SRPM) with inverse temperature $%
\beta =1$ for orthogonal model and $\beta =2$ for unitary. These critical
level statistics are argued to be universal by comparing results from
systems both with and without total $S_{z}$ conservation. We also point out
that these critical distributions can emerge from the spectrum of a Poisson
ensemble, which indicates the thermal-MBL transition point is more affected
by the MBL phase rather than thermal phase.
\end{abstract}

\maketitle

\section{Introduction}

\label{sec1}

\bigskip The non-equilibrium phases of matter in isolated quantum systems is
a focus of modern condensed matter physics, it is now well-established the
existence of two generic phases: a thermal phase and a many-body localized
(MBL) phase\cite{Gornyi2005,Basko2006}. Physically, a thermal phase is
ergodic with extended and featureless eigenstate wavefunctions, which
results in a correlated eigenvalue spectrum with level repulsion. In
contrary, in MBL phase localization persists in the presence of weak
interactions. Modern understanding about these two phases relies on quantum
entanglement. In thermal phase, the system acts as the heat bath for its
subsystem, hence the entanglement is extensive and exhibits ballistic
(linear in time) spreading after quantum quench. In contrast, the absence of
thermalization in MBL phase leads to small (area-law) entanglement and slow
(logarithmic) entanglement spreading. The qualitative difference in the
scaling of quantum entanglement and its dynamics after quantum quench are
widely used in the study of thermal-MBL transition\cite%
{Kjall2014,Geraedts2017,Yang2015,Serbyn16,Gray2017,Maksym2015,Kim,Bardarson,Abanin}%
.

More traditionally, the thermal phase and MBL phase is distinguished by
their eigenvalue statistics, whose theoretical foundation is provided by the
random matrix theory (RMT)\cite{Mehta,Haake2001}. RMT is a powerful
mathematical tool that describes the universal properties of a complex
system that depend only on its symmetry while independent of microscopic
details. Specifically, the Gaussian orthogonal ensemble (GOE) describes
systems with spin rotational and time reversal symmetry; the Gaussian
unitary ensemble (GUE) corresponds to those with spin rotational invariance
and broken time reversal symmetry; and Gaussian symplectic ensemble (GSE)
refers to systems which conserve time reversal symmetry while break spin
rotational invariance. It is well established that in the thermal phase with
correlated eigenvalues, the distribution of nearest level spacings $\left\{
s_{i}=E_{i+1}-E_{i}\right\} $ will follow a Wigner-Dyson distribution with
Dyson index $\beta =1,2,4$ for GOE,GUE,GSE respectively. On the other hand,
in MBL phase with uncorrelated eigenvalues, $P\left( s\right) $ is expected
to follow Poison distribution. The difference in the level spacing
distribution is also widely-used in the study of MBL\ systems\cite%
{Oganesyan,Avishai2002,Regnault16,Regnault162,Huse1,Huse2,Huse3,Luitz}.

\bigskip Compared to the properties of each phase, the nature of the
thermal-MBL transition is much less understood. Many works on
one-dimensional MBL system indicate the existence of Griffiths regime near
the transition point, where the system becomes an inhomogeneous mixture of
locally thermal and localized regions. Consequently,\ the system's dynamics
become anomalously slow and eigenstates exhibits multifractality. However,
this regime is not free of uncertainties, and a unified theory has not been established by now\cite%
{Alet,Gopalakrishnan,Agarwal2015,Agarwal2017,Luitz2017,Mace2019}.

Despite the lack of understanding about the thermal-MBL transition, there
are a number of effective models proposed for the critical level statistics
at the transition point. For example, the Rosenzweig-Porter model\cite%
{Shukla}, mean field plasma model\cite{Serbyn}, short-range plasma models
(SRPM)\cite{SRPM} and its generalization -- the weighed SRPM\cite{Sierant19}%
, Gaussian $\beta $ ensemble\cite{Buijsman} and the generalized $\beta -h$
model\cite{Sierant20}, and others\cite{Ndawana,Ray}. In this work, we will
focus on the SRPM, whose formal definition will be given in Sec.~\ref{sec2}.
Historically, SRPM is introduced as a RMT model that holds the semi-Poisson
level statistics, which is an intermediate statistics between GOE and
Poisson that close to the one found numerically at the critical point of
Anderson metal-insulator transition\cite{Shklovskii}. As for the MBL
transition, SRPM with inverse temperature $\beta =1$ has been shown to
describe the nearest level spacing distribution at critical region well,
while its effectiveness in describing long-range level correlations is
debated\cite{Sierant19}. In this work, we will study the higher-order level
spacings that incorporate level correlations on longer ranges, and show the
SRPM is indeed a good effective model for the critical region, at least when
level correlations on moderate ranges are concerned.

Besides, current works on the thermal-MBL transition are mostly dealing with
orthogonal systems, whose corresponding RMT description is GOE to Poisson.
It is natural to ask what's the critical spacing distributions in a unitary
system, and what's the corresponding effective model. Given the RMT
description for MBL transition in a unitary system is GUE to Poisson, a
natural candidate for the effective model would be the SRPM with inverse
temperature $\beta =2$. It is the second purpose of current work to verify
this guess.

In this paper, we study the level statistics in the thermal-MBL transition
region of 1D random spin systems, our analysis relies solely on the energy
spectrum. By using the inter-sample randomness as the ``order parameter'',
we quantitatively locate the transition points, which are in well-agreement
with previous results based on eigenstate properties. We further count the $%
n $-th order level correlations in the transition region up to $n=4$, and
verify they fit well with those of SRPM with inverse temperature $\beta =1$
for orthogonal model and $\beta =2$ for unitary, and these critical
behaviors are expected to be universal by comparing results from models both
with and without total $S_{z}$ conservation. We also discuss how the SRPM
can emerge from the eigenvalue spectrum of the MBL phase, indicating the
thermal-MBL transition point is more affected by the MBL phase rather than
thermal phase.

This paper is organized as follows. In Sec.~\ref{sec2} we introduce the spin
model and SRPM. In Sec.~\ref{sec3} we focus on the orthogonal models, and
unitary models are studied in Sec.~\ref{sec4}. Conclusion and discussion
come in Sec.~\ref{sec5}.

\section{Model and Method}

\label{sec2}

We will study the \textquotedblleft standard model\textquotedblright\ for
MBL physics, i.e., the anti-ferromagnetic Heisenberg model with random
external fields, whose Hamiltonian is%
\begin{equation}
H=\sum_{i=1}^{L}\mathbf{S}_{i}\cdot \mathbf{S}_{i+1}+\sum_{i=1}^{L}\sum_{%
\alpha =x,y,z}h^{\alpha }\varepsilon _{i}^{\alpha }S_{i}^{\alpha }\text{,}
\label{equ:H}
\end{equation}%
where $\mathbf{S}_{i}$ is spin-$1/2$ operators. The anti-ferromagnetic
coupling strength is set to be $1$, and periodic boundary condition is
assumed in the Heisenberg term. The $\varepsilon _{i}^{\alpha }$s are random
variables within range $\left[ -1,1\right] $, and $h^{\alpha }$ is referred
as the randomness strength. This Hamiltonian's property depends on the
external fields: when they are non-zero in only one or two spin directions,
the model is orthogonal; while when all of them are non-zero, the model is
unitary. In all cases, the system will undergo a thermal-MBL transition with
increasing randomness, and the corresponding RMT description is GOE (GUE) to
Poisson in the orthogonal (unitary) case.

To describe the level statistics, we choose to study the distributions of
the nearest gap ratios, whose definition is%
\begin{equation}
r_{i}=\frac{s_{i+1}}{s_{i}}\equiv \frac{E_{i+2}-E_{i+1}}{E_{i+1}-E_{i}}\text{%
.}
\end{equation}%
Compared to the more traditional quantity of level spacings $\left\{
s_{i}=E_{i+1}-E_{i}\right\} $, the gap ratios $\left\{ r_{i}\right\} $ have
two major advantages: (i) unlike level spacings, $P\left( r\right) $ is
independent of density of states (DOS), hence requires no unfolding
procedure, which is non-unique and may raise subtle misleading signatures
when studying the long-range level correlations in some systems\cite%
{Gomez2002}; (ii) counting $P\left( s\right) $ requires an additional
normalization for $\langle s\rangle $, while counting $P\left( r\right) $
does not. Actually, the mean value $\langle r\rangle $ can be a measure
to distinguish phases, as has been adopted in many recent works.

The analytical form of $P\left( r\right) $ for the thermal phase has been
derived in Ref.~[\onlinecite{Atas}] using a Wigner-like surmise, which gives%
\begin{equation}
P\left( \beta ,r\right) =Z_{\beta }\frac{\left( r+r^{2}\right) ^{\beta }}{%
\left( 1+r+r^{2}\right) ^{1+3\beta /2}}\text{,}  \label{equ:1}
\end{equation}%
where the Dyson index $\beta =1,2,4$ stands for GOE,GUE,GSE respectively,
and $Z_{\beta }$ is a normalization factor determined by $\int_{0}^{\infty
}P\left( \beta ,r\right) dr=1$. The gap ratio can be generalized to higher
order to describe level correlations on longer ranges, whose definition is%
\begin{equation}
r_{i}^{\left( n\right) }=\frac{E_{i+2n}-E_{i+n}}{E_{i+n}-E_{i}}\text{,}
\end{equation}%
and the corresponding distribution is\cite{Tekur,Rao20}%
\begin{eqnarray}
P\left( \beta ,r^{\left( n\right) }=r\right) &=&P\left( \gamma ,r\right)
\text{,} \\
\gamma &=&\frac{n\left( n+1\right) }{2}\beta +n-1\text{.}  \notag
\end{eqnarray}%
On the other hand, for the MBL phase with uncorrelated energy spectrum, we
have\cite{Atas2,Rao202}%
\begin{equation}
P\left( r^{\left( n\right) }=r\right) =\frac{r^{n-1}}{\left( 1+r\right) ^{2n}%
}\text{.}  \label{equ:poi}
\end{equation}

As for the spectral statistics at the thermal-MBL transition region, a
number of effective models have been proposed, and in this work we will
focus on the short-range plasma model (SRPM). The SRPM describes the
eigenvalues of a random matrix ensemble as an ensemble of one-dimensional
system of classical particles with two-body repulsive interactions, whose
distribution can be written into a canonical ensemble form%
\begin{eqnarray}
P_{\beta }\left( \left\{ E_{i}\right\} \right) &=&Z_{\beta }^{-1}e^{-\beta
H\left( \left\{ E_{i}\right\} \right) }\text{,} \\
H\left( \left\{ E_{i}\right\} \right) &=&\sum_{i}U\left( E_{i}\right)
+\sum_{\left\vert i-j\right\vert \leq k}V\left( \left\vert
E_{i}-E_{j}\right\vert \right) \text{,}  \label{equ:SRPM}
\end{eqnarray}%
where $U\left( E_{i}\right) \varpropto E_{i}^{2}$ is the trapping potential
and the Dyson index $\beta $ is interpreted as the inverse temperature. The
two-body interaction takes the logarithmic form $V\left( x\right) =-\log
\left\vert x\right\vert $, and $k$ is the interaction range. It's easy to
see the $k\rightarrow \infty $ limit corresponds to the standard Gaussian
ensembles for thermal phase; while in $k\rightarrow 0$ limit no interaction
is present, which corresponds to the Poisson ensemble with no level
correlation; the thermal-MBL transition is thus reflected by the evolution
of the interaction range $k$. Unlike the mean-field plasma model, which is
also suggested to describe the critical spectral statistics\cite{Serbyn}, it
is the interaction range rather than the interaction form that changes
during the thermal-MBL transition.

One major advantage of SRPM is that it is exactly solvable, and the general
form of $n$-th order level spacing distribution has been derived in Ref.~[%
\onlinecite{SRPM}]. Notably, for the simplest case with $\beta =k=1$, the
nearest level spacings $\left\{ s_{i}\right\} $ follows the semi-Poisson
distribution, which is close to the one found numerically at the MBL
transition region in an orthogonal spin model\cite{Huse2,Regnault16,Serbyn}.
In this work, we will proceed to study the higher-order gap ratios $\left\{
r_{i}^{\left( n\right) }\right\} $ in SRPM that incorporate level
correlations on longer ranges. Unlike the more traditional quantities such
as number variance $\Sigma ^{2}$, the higher-order gap ratios are
numerically easier to obtain and require no unfolding procedure hence avoid
the potential ambiguity raised by concrete unfolding strategy\cite{Gomez2002}%
.

First of all, we need to get the expression of $P\left( r^{\left( n\right)
}\right) $ for the SRPM, which is not an easy task since a Wigner-like
surmise is not applicable due to the limited interaction range in Eq.~(\ref%
{equ:SRPM}). However, we can make use of an elegant correspondence between
the SRPM and the ``reduced energy
spectrum'' of Poisson ensemble, whose idea goes as follows.

Formally, a $r$-th order
reduced energy spectrum $\left\{ E_{i}^{\left( r\right) }\right\} $ is
comprised of every $\left( r+1\right) $-th level of the original spectrum $%
\left\{ E_{i}\right\} $, which is mathematically achieved by tracing out
every $r$ levels in between. This construction is very similar to that of
the reduced density matrix where we trace out the degrees of freedom in a
subsystem, hence we suggest to call $\left\{ E_{i}^{\left( r\right)
}\right\} $ the \textquotedblleft reduced energy spectrum\textquotedblright
\cite{Rao202}. It is proved in Ref.~[\onlinecite{Daisy}] that the energy
spectrum of SRPM with $k=1$ and inverse temperature $\beta $ has the same
structure as the $\beta $-th order reduced energy spectrum of a Poisson
ensemble (which is named \textquotedblleft Daisy model\textquotedblright\ by
the authors). By this mapping, the $n$-th eigenvalue in the SRPM with
inverse temperature $\beta $ becomes the $n\left( \beta +1\right) $-th level
in the Poisson ensemble, and the $n$-th order gap ratio in the former is
mapped to the $n\left( \beta +1\right) $-th order counterpart in the latter,
whose distribution is then easily written down according to Eq.~(\ref%
{equ:poi}), that is%
\begin{equation}
P\left( \beta ,r^{\left( n\right) }=r\right) =\frac{r^{\gamma -1}}{\left(
1+r\right) ^{2\gamma }}\text{, }\gamma =n\left( \beta +1\right) \text{.}
\label{equ:HighOrder}
\end{equation}%
In the next sections, we will use Eq.~(\ref{equ:HighOrder}) with $\beta =1$ (%
$2$) to fit the critical level statistics in orthogonal (unitary) model.
Besides, by comparing results from models both with and without total $S_{Z}$
conservation, we argue that this effective model is universal that
independent of microscopic details.

\section{Orthogonal Models}

\label{sec3}

\bigskip We start by studying the orthogonal models in Eq.~(\ref{equ:H}). We
first consider the case that $h^{x}=h^{z}=h\neq 0$ and $h^{y}=0$. This
choice breaks total $S_{Z}$ conservation and makes the eigenstates in
thermal phase fully featureless, hence is less affected by finite-size
effect. The MBL transition point is, according to previous studies\cite%
{Regnault16,Regnault162}, $h_{c}\simeq 3$. Note that, although the pure
Heisenberg chain has different energy spectrums in systems of even and odd
lengths, the difference is wiped out by the random external fields. In this
work, we study up to $L=13$ system with Hilbert space dimension $%
N=2^{13}=8192$.

To get an intuitive picture of the gap ratio's evolution, we numerically
simulate Eq.~(\ref{equ:H}) in a $L=13$ system in the randomness range $h\in %
\left[ 1,5\right] $, with $500$ samples taken at each randomness strength.
For each energy spectrum sample, we select $25\%$ eigenvalues in the middle
to determine $P\left( r\right) $, and the results are displayed in Fig.~\ref%
{fig:Orth}(a). As we can see, when the randomness is small ($h=1$), $P\left(
r\right) $ meets perfectly with the prediction for GOE; when randomness
increases, $P\left( r\right) $ starts to deform, and finally reach to the
Poisson distribution for MBL\ phase ($h=5$). We note the fittings for $h=5$
has minor deviations from ideal Poisson, this is due to finite-size effect,
since in a finite system there will always remain exponentially decaying but
finite correlations between eigenstates.

\begin{figure}[t]
\centering
\includegraphics[width=\columnwidth]{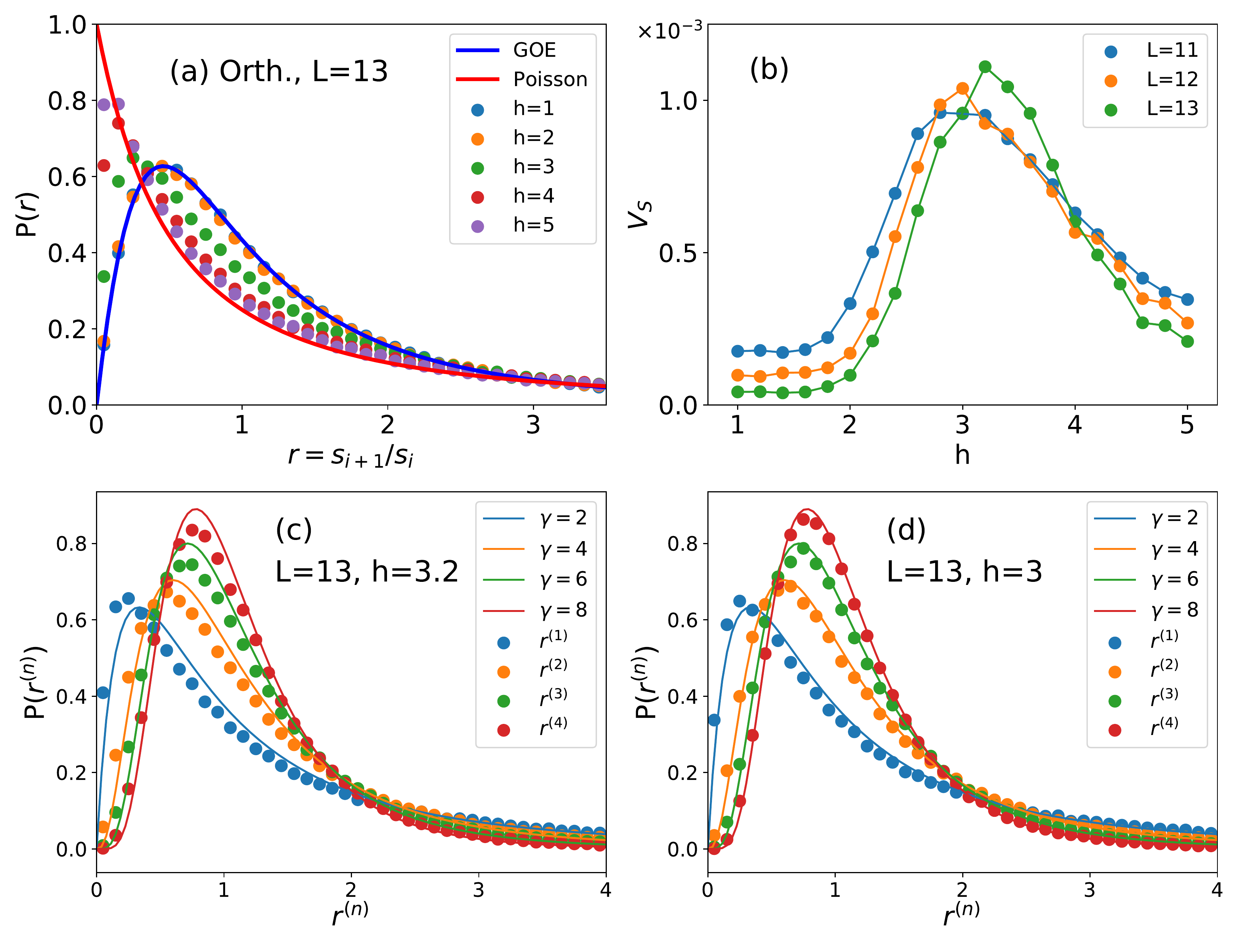}
\caption{(a) The evolution of gap ratio distribution in an $L=13$ orthogonal system,
$P(r)$ evolves from GOE to Poisson when the system is evolved from thermal
($h=1$) to MBL phase ($h=5$). (b) The evolution of inter-sample variance $V_S$ as a function of $h$, in systems with different sizes. Both the peak position and value of $V_S$ are larger in larger system, indicating a larger Giffiths regime.
(c) $P\left(r^{(n)}\right)$ at the estimated transition point
$h=3.2$ in an $L=13$ system, non-negligible deviations from SRPM are found due to residue correlations raised
by finite-size effect. (d) $P\left(r^{(n)}\right)$ at $L=13$ and $h=3$, perfect matches with SRPM ($\beta=1$) are observed.
} \label{fig:Orth}
\end{figure}

From the evolution of $P\left( r\right) $ in Fig.~\ref{fig:Orth}(a), we can
have a qualitative estimation about the location of MBL transition point. To
be specific, take a closer look at $P\left( r\right) $ at $h=3$ (the green
dots in Fig.~\ref{fig:Orth}(a)), we see it lies roughly at the middle
between GOE and Poisson, which indicates $h=3$ is close to the transition
point.

To more quantitatively locate the transition point, we adopt a variant
definition of gap ratio, which is%
\begin{equation}
t_{i}=\frac{\min \left\{ s_{i+1},s_{i}\right\} }{\max \left\{
s_{i+1},s_{i}\right\} }
\end{equation}%
where $s_{i}=E_{i+1}-E_{i}$ is the $i$-th energy gap. This is actually the
original definition of gap ratio introduced by Oganesyan and Huse\cite%
{Oganesyan}. Compared to $r_{i}$, $t_{i}$ takes values in the range $(0,1]$,
and their distributions are related by $P\left( t\right) =2P\left( r\right)
\Theta \left( 1-r\right) $\cite{Atas}. The mean value of gap ratio $%
\overline{t}$ can be easily calculated from Eq.~(\ref{equ:1}), namely $%
\overline{t}_{\text{GOE}}=0.536$, $\overline{t}_{\text{GUE}}=0.603$ and $%
\overline{t}_{\text{Poisson}}=0.386$. Technically, the calculation of $%
\overline{t}$ has two steps: first we calculate the mean gap ratio value in
\emph{one sample}, which gives $t_{S}=\langle t_{i}\rangle _{\text{samp}}$,
then we average $t_{S}$ over an ensemble of samples to get $\overline{t}%
=\langle r_{S}\rangle _{\text{en}}$. These two steps give two types of
variance, the first one is $V_{S}=\langle t_{S}^{2}-\overline{t}^{2}\rangle
_{\text{en}}$, i.e. the variance of sample-averaged gap ratio over ensemble,
which measures the \emph{inter-sample randomness}; another one is $%
V_{I}=\langle v_{I}\rangle _{\text{en}}$ where $v_{I}=\langle
t_{i}^{2}-t_{S}^{2}\rangle _{\text{samp}}$, which is the ensemble-averaged
gap ratio variance and measures the intrinsic \emph{intra-sample randomness}%
. In a system driven by pure random disorder (that is, opposite to the ones
induced by quasi-periodic potential\cite{RD}), the distribution of $t_{S}$
near the transition region will exhibit strong deviation from a Gaussian
type -- a manifestation of Griffiths region -- which results in a peak value
of $V_{S}$ at the transition point\cite{Sierant19}. Therefore, for our model
Eq.~(\ref{equ:H}), we can calculate the evolution of $V_{S}$ to locate the
transition point.

Strictly speaking, the transition point identified by $V_{S}$ and other
quantities based on quantum entanglement may not always coincide in a \emph{%
finite system}, meanwhile, $V_{S}$ in essence describes a \emph{qualitative}
structural change in the energy spectrum, hence is more suitable for our
purpose to study the critical level statistics.

In Fig.~\ref{fig:Orth}(b) we draw the evolution of $V_{S}$ in systems with
different lengths, where the number of samples are $10000,2000,500$ for $%
L=11,12,13$, respectively. In all cases, expected peaks of $V_{S}$ appear. We see that, in
general, both the detected transition point and peak value $V_{S}$ are
larger in larger system, which indicates a larger Griffiths regime, in
consistence with the results in orthogonal model with $S_{Z}$ conservation\cite%
{Sierant19}.

Now we are ready to count the level statistics at the transition region. In
a finite system, what we observe is always a combination of universal part
and non-universal (model dependent) part, we therefore choose the largest
system we can reach to minimize the finite-size effects. That is, $L=13$ for
systems without $S_{Z}$ conservation, and $L=16$ for those with $S_{Z}$
conservation. As for the present model with $L=13$, the detected transition
point is, according to Fig.~\ref{fig:Orth}(b), $h_{c}\simeq 3.2\pm 0.2$.

First we take out the samples at the identified transition point $h=3.2$,
and determine the corresponding gap ratio distributions $P\left( r^{\left(
n\right) }\right) $ up to $n=4$, the results are displayed in Fig.~\ref%
{fig:Orth}(c), where the reference curves are the ones for SRPM in Eq.~(\ref%
{equ:HighOrder}) with $\beta =1$. As we can see, the fittings have
non-negligible deviations. We attribute this to the finite system size we
are studying. That is, in a finite system, the eigenstates even in the MBL
phase remain an exponentially decaying but finite correlations, hence the
randomness required to drive the phase transition is slightly larger than it
really needs in thermodynamic limit, which is in agreement with our analysis
for Fig.~\ref{fig:Orth}(a). Therefore, the true critical level statistics is
expected to occur in a point slightly smaller than $h=3.2$. To this end, we
take out the samples from $h=3$ and count the corresponding $P\left(
r^{\left( n\right) }\right) $, the results are in Fig.~\ref{fig:Orth}(d). As
can be seen, they fit quite well with the SRPM, confirming the SRPM is
indeed a good effective model, at least when level correlations on moderate
ranges (up to $9$ levels) are concerned. To be complete, we have checked the
same situations happen in the $L=12$ system.

To show this critical distribution is universal, we consider another
orthogonal model, that is, the one with $h^{z}=h\neq 0,h^{x}=h^{y}=0$ in
Eq.~(\ref{equ:H}). This is actually the one most widely studied in the
literature since it preserves total $S_{z}$, and allows one to reach to
larger system size by focusing on one sector, which is commonly chosen to be
the one with $S_{Z}^{T}=0$. Technically, this also requires the number of
spins $L$ to be even. However, eigenstates in this sector share one common
feature, i.e. $S_{Z}^{T}=0$, which violates the featureless property of a
fully thermalized state. Therefore, the eigenstates in this sector is easier
to be localized, which results in a large large finite-size effect, and the
estimated transition point is much less smaller than the interpolated value
in thermodynamic limit. Actually, it's widely accepted the transition point
is around $h_{c}\simeq 3.6$ for the middle part of energy spectrum, while in
a finite system, say $L=16$, the detected transition point is shifted to $%
h_{c}\simeq 2.7$\cite{Sierant19}.

\begin{figure*}[htbp]
\centering
\includegraphics[width=1.8\columnwidth]{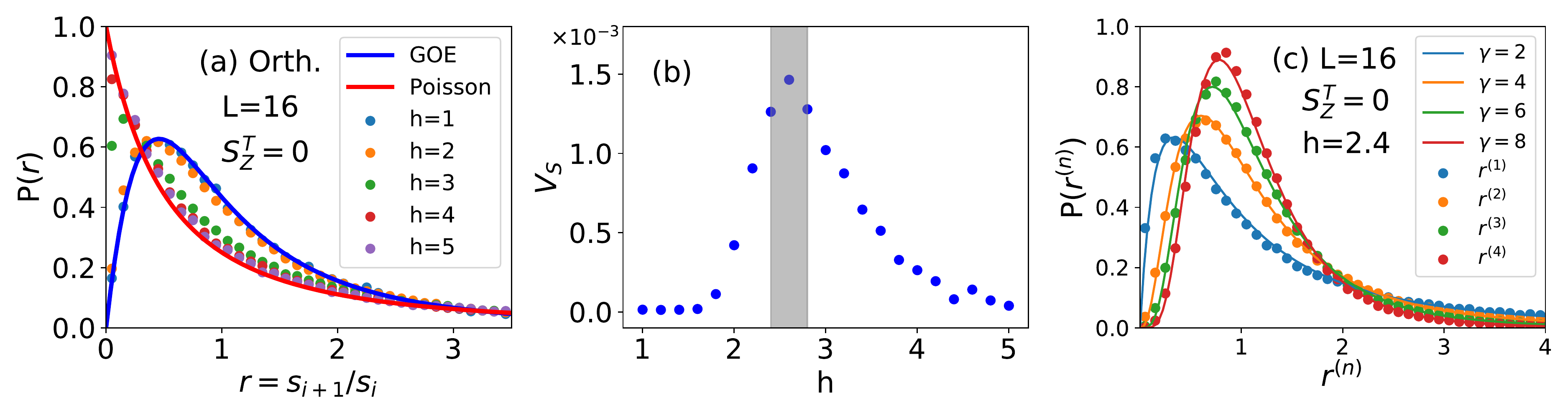}
\caption{(a) The evolution of gap ratio distribution $P(r)$ in the $S_Z^T=0$
sector of the $L=16$ orthogonal model, an expected GOE-Poisson transition is
found. (b) The evolution of $V_S$, which indicates the transition point is $%
h_c=2.6\pm0.2$. (c) $P\left(r^{(n)}\right)$ at $h=2.4$, a perfect match with
SRPM ($\protect\beta=1$) is observed, indicating this effective model is
universal.}
\label{fig:Sym_orth}
\end{figure*}

\bigskip In our study, we take the system size to be $L=16$ and focus on the
$S_{Z}^{T}=0$ sector, whose Hilbert space's dimension is $N=C_{16}^{8}=\frac{%
16!}{8! 8!}=12870$. Like before, we first present a qualitative picture for
the gap ratio's evolution in the range $h\in \left[ 1,5\right] $, with $500$
samples taken at each point, the results are displayed in Fig.~\ref%
{fig:Sym_orth}(a), we see a GOE-Poisson evolution as expected. Then we
numerically determine the evolution of inter-sample randomness $V_{S}$,
which is presented in Fig.~\ref{fig:Sym_orth}(b). The observed peak
indicates a transition at $h=2.6\pm 0.2$, in well accordance with the
previous studies in this system size. Next we consider the critical
statistics. With the same reason as for previous model, we take the samples
from $h=2.4$, slightly smaller than the estimated one, and the corresponding
$P\left( r^{\left( n\right) }\right) $ are displayed in Fig.~\ref%
{fig:Sym_orth}(c). As can be seen, they fit quite good with the prediction
of SRPM with $\beta =1$.

Up to now, we have confirmed the SRPM with $k=\beta =1$ is a quite good
effective model for the critical spectral statistics in an orthogonal model,
not only for nearest-neighbor gap ratios, but also for several higher-order
ones that describe level correlations on longer ranges, and this model is
expected to be universal that independent of microscopic details. In the
next section, we will proceed to study the unitary model.

\section{Unitary Models}

\label{sec4}

\begin{figure*}[t]
\centering
\includegraphics[width=2\columnwidth]{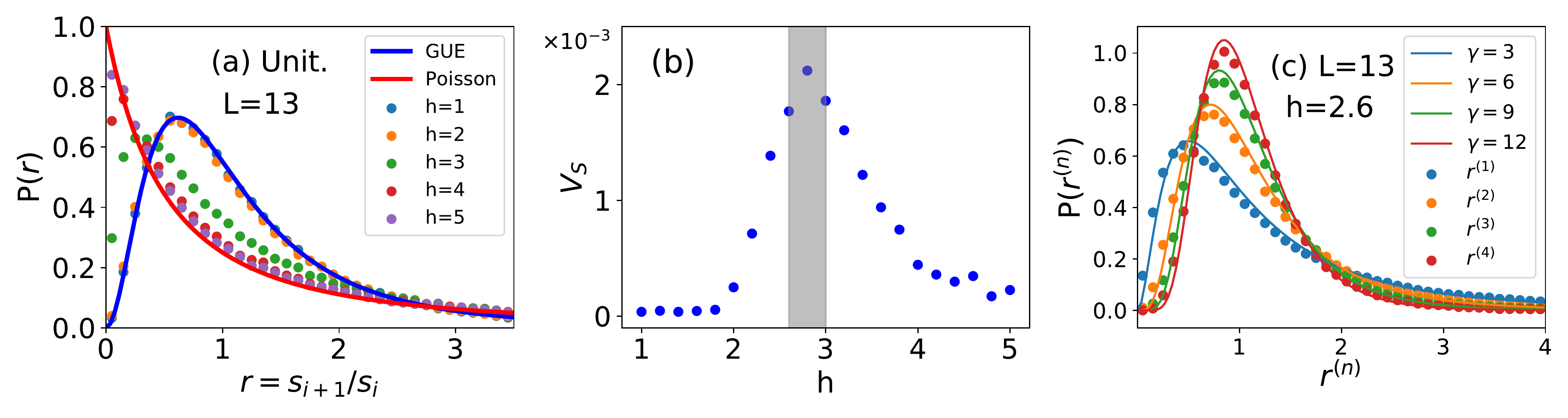}
\caption{(a) The evolution of $P(r)$ in an $L=13$ unitary system, a
GUE-Poisson transition is observed as expected. (b) The evolution of
inter-sample randomness $V_S$, and the shaded area indicates the transition
region $h_c=2.8\pm0.2$. (c) $P\left(r^{(n)}\right)$ at $h=2.6$, a good match
with SRPM ($\protect\beta=2$) is observed.}
\label{fig:Unit}
\end{figure*}

\bigskip Now we study the critical level statistics in unitary models, we
will show it is well described by SRPM with $k=1$ and $\beta =2$. Like
before, we first consider the case without $S_{Z}^{T}$ conservation, that
is, the model Eq.~(\ref{equ:H}) with $h_{x}=h_{y}=h_{z}=h\neq 0$. Likewise,
we work on an $L=13$ system, the qualitative evolution of $P\left( r\right) $
is given in Fig.~\ref{fig:Unit}(a), a GUE-Poisson evolution is observed when
increasing randomness as expected. From the evolution, we can qualitatively
see the transition point lies at somewhere between $h=2$ and $h=3$. Next, we
calculate the evolution of inter-sample randomness $V_{S}$, the result is
given in Fig.~\ref{fig:Unit}(b). We see an expected peak indicating the
transition point is $h_{c}\simeq 2.8\pm 0.2$, close to $h_{c}\simeq 2.5$ got
by previous studies\cite{Regnault16,Regnault162}.

Next, we are considering the critical level statistics. With the same reason
as in previous section, we take a point slightly left to the estimated one,
that is $h=2.6$, and the corresponding $P\left( r^{\left( n\right) }\right) $
are presented in Fig.~\ref{fig:Unit}(c). As can been seen, they fit very
well with the predictions of SRPM with $k=1$ and $\beta =2$, which provides
a strong evidence that the SRPM is a good effective model.

To further show this critical behavior in the unitary model is also
universal, we study a unitary spin model with total $S_{Z}$ conservation,
which is constructed by adding a time-reversal breaking next-nearest
neighboring interaction term to the Heisenberg model, the Hamiltonian then
reads%
\begin{equation}
H=\sum_{i=1}^{L}\left[ \mathbf{S}_{i}\cdot \mathbf{S}_{i+1}+J\mathbf{S}%
_{i}\cdot \left( \mathbf{S}_{i+1}\times \mathbf{S}_{i+2}\right) \right]
+h\sum_{i=1}^{L}\varepsilon _{i}^{z}S_{i}^{z}\text{.}
\end{equation}%
This model was introduced to generate the GUE statistics in Ref.~[%
\onlinecite{Avishai2002}], it was also pointed out the level statistics
almost immediately changes from GOE to GUE even when $J$ is as small as $%
0.01 $\cite{Avishai2002}. The thermal-MBL transition point in this model
certainly depends on $J$: in general, the larger $J$, the larger $h_{c}$
will be. In this work, we choose $J=0.2$ without loss of generality, and
focusing on the $S_{Z}^{T}=0$ sector in an $L=16$ system.

\begin{figure*}[t]
\centering
\includegraphics[width=2\columnwidth]{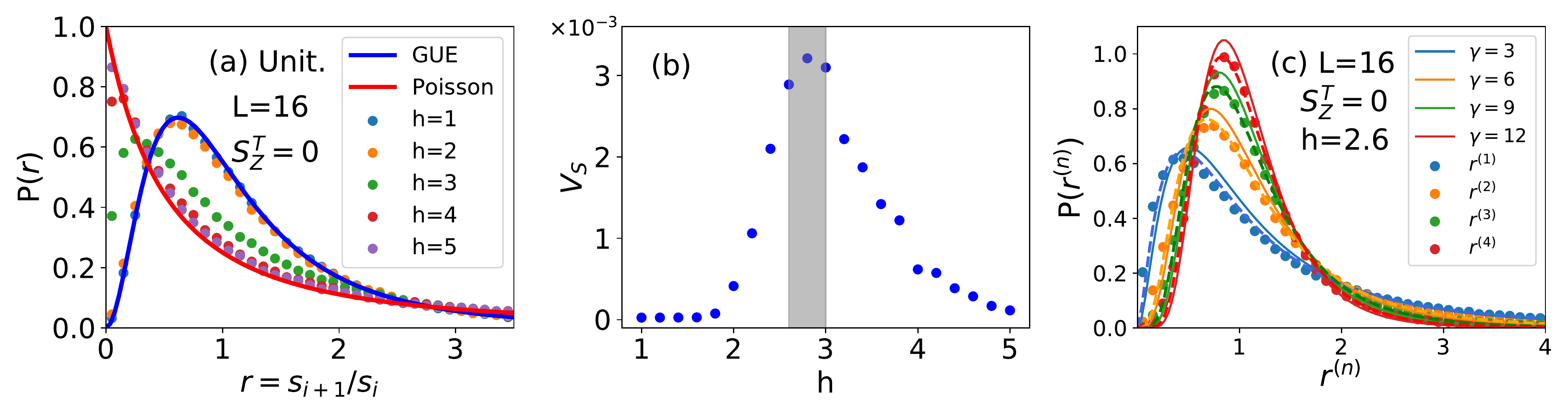}
\caption{(a) The evolution of $P(r)$ in the $S_Z^T=0$ sector of the $L=16$
unitary model with the next-nearest neighboring interaction strength $J=0.2$%
, an expected GUE-Poisson transition is found. (b) The evolution of $V_S$,
which indicates the transition region is $h_c=2.8\pm0.2$. (c) $%
P\left(r^{(n)}\right) $ at $h=2.6$, minor deviations from SRPM with $\protect%
\beta=2$ are observed, which may be attributed to the large finite-size
effect induced by the next-nearest neighboring interaction that destroys the
integrability of pure Hamiltonian. The dotted lines: SRPM with $\protect\beta%
=1.7$.}
\label{fig:Sym_unit}
\end{figure*}

The qualitative evolution of $P\left( r\right) $ is given in Fig.~\ref%
{fig:Sym_unit}(a), a GUE-Poisson evolution when increasing randomness $h$ is
observed as expected. Next, we calculate the evolution of inter-sample
randomness $V_{S}$, the result is presented in Fig.~\ref{fig:Sym_unit}(b).
We see an expected peak indicating the transition point is $h_{c}\simeq
2.8\pm 0.2$. Interestingly, this coincides with the one in Fig.~\ref%
{fig:Unit}(b), which is purely accidental for the $J=0.2$ we choose.
Actually, the values of $V_{S}$ in Fig.~\ref{fig:Sym_unit}(b) are much
larger than those in Fig.~\ref{fig:Unit}(b), which means the inter-sample
randomness is generally larger in this model, hence the finite-size effect
is expected to be more serious.

Next, we are considering the critical statistics. Like before, we take a
point slightly smaller than the estimated one, that is $h=2.6$, and the
corresponding $P\left( r^{\left( n\right) }\right) $ are presented in Fig.~%
\ref{fig:Sym_unit}(c). As can been seen, they qualitatively meets the
predictions of SRPM with $k=1$ and $\beta =2$, but the deviations are larger
than those in Fig.~\ref{fig:Unit}(c). We attribute this to the next-nearest
interaction term that destroys the integrability of the pure model, which
strengthens the thermal phase in the random model and results in a more
serious finite-size effect, in accordance with our analysis about Fig.~\ref%
{fig:Sym_unit}(b). In fact, if we artificially allow the inverse temperature
to be a fraction, we find the $P\left( r^{\left( n\right) }\right) $ in Fig.~%
\ref{fig:Sym_unit}(c) can be well fitted into $\beta =1.7$ (the dotted lines
in Fig.~\ref{fig:Sym_unit}(c)), which is close to the expected value $\beta
=2$.

To conclude, we have shown the spectral statistics in the transition region
of a unitary model without $S_{Z}^{T}$ conservation is well described by the
SRPM with $k=1$ and $\beta =2$, which is a natural extension of the one for
the orthogonal system. The results from unitary model with $S_{Z}^{T}$
conservation suggests this critical behavior is also universal for unitary
model, although the deviations are slightly larger. We suggest a future work
on larger system to confirm this conclusion.

\section{Conclusion and Discussion}

\label{sec5}

We have studied the thermal-MBL transition in both orthogonal and unitary
models in random spin systems. By using the inter-sample randomness as the
``order parameter'', we successfully located the transition points, which
are in well agreement with previous studies. We then determine the $n$-th
order gap ratio distributions up to $n=4$ at the critical region, and
confirm they fit well with the short-range plasma model (SPRM) with inverse
temperature $\beta =1$ for orthogonal model and $\beta =2$ for unitary.
Based on results from models both with and without $S_{Z}^{T}$ conservation,
we argue these critical behaviors are universal that independent of
microscopic details.

It is worth noting that the level statistics right at the transition points
detected by $V_{S}$ show systematic deviations from SRPM in all cases
studied, this is due to the finite size effect. To be precise, in a finite
system, there will always remain exponentially decaying but finite level
correlations even deep in the MBL phase, as can be seen from the fitting
results for MBL phases in Fig.1(a),2(a),3(a) and 4(a). Therefore, the
randomness strength to drive the phase transition will be larger to
compensate for these residue level correlations. Consequently, the true
critical statistics will appear slightly left to the detected transition
point. The deviations in the unitary model with $S_{Z}^{T}$ conservation
are larger than the rest models, which can be attributed to the
neat-nearest neighboring interactions. That is, the next-nearest neighboring
term breaks the integrability of the clean system and stabilizes the thermal
phase in disordered system, which results in larger residue level
correlations in a finite system. After all, in all cases, what we observe is
a combination of universal critical level statistics and non-universal
(model-dependent) finite-size results, for which a detailed quantification
will require a systematic finite-size scaling study, and is left for a
future work.

It would be beneficial to compare the SRPM with other proposed effective
models for the transition region, first of which is the mean-field plasma
model (MFPM), which is proposed by Serbyn and Moore by mapping the
thermal-MBL transition into a random walk process in the Hilbert space\cite%
{Serbyn}. Mathematically, both SRPM and MFPM describe the energy levels of a
random matrix ensemble as an ensemble of 1D classical particles, however in
SRPM the interaction form stays unchanged and interaction range is
responsible for the thermal-MBL transition, while the inverse is true for
the MFPM. Meanwhile, both models hold the semi-Poisson distribution for the
nearest level spacings, hence our results are not controversial to those in
Ref.~[\onlinecite{Serbyn}]. In this study, we proceed to consider the
higher-order level correlations, and find good support for the SRPM. In
fact, our results suggest the form of local interaction between energy
levels stays logarithmic during the phase transition, and the change in
interaction range can be revealed by the high-order level correlations.
Another proposed effective model is the Gaussian $\beta $ ensemble, which
has the same structure as the GOE,GUE,GSE but the Dyson index $\beta $ takes
value in $\left( 0,\infty \right) $. In Ref.~[\onlinecite{Buijsman}] the
authors showed the Gaussian ensemble with non-integer $\beta $ can describe
the lowest-order gap ratio distribution across the thermal-MBL transition
quite well but the fittings for higher-order ones have large deviations.
This also suggests the form of interaction between levels stays logarithmic
but the interaction range changes during phase transition, hence is also
consistent with our results.

Another interesting fact to notice is that the SRPM can emerge from the
Poisson ensemble, that is, the SRPM with $k=1$ and inverse temperature $%
\beta $ has the same structure as the $\beta $-th order reduced energy
spectrum of a Poisson ensemble\cite{Daisy} (which is called ``Daisy model''
by the authors). This indicates the universal lower-order spectral
statistics at the transition region are secretly hidden in the eigenvalue
spectrum of the MBL phase, for which a full physical understanding is
lacking by now. However, this at least indicates the thermal-MBL transition
point is more affected by the MBL phase rather than the thermal phase, a
fact that has already been noticed by previous studies based on
eigenfunction properties\cite{Huse2,Serbyn,Sierant19} and now appears again
by means of the reduced energy spectrum. On the other hand, in Ref.~[%
\onlinecite{Daisy}] the authors declare the absence of a dynamical system
that corresponds to the ``Daisy model'' with inverse temperature $\beta >1$,
our work thus suggests the thermal-MBL transition point in unitary system is
a natural candidate for $\beta =2$.

Last but not least, the SRPM was debated for its effectiveness in describing
long range level correlations at the MBL transition\cite{Sierant19}, e.g.
through the\ number variance $\Sigma ^{2}$. Unfortunately our attempts to
fit $\Sigma ^{2}$ do not give conclusive results. This may partially due to
the intrinsic sensitive dependence of $\Sigma ^{2}$ on concrete unfolding
procedure\cite{Gomez2002}, and also may results from the limited system size
we can reach. Nevertheless, our results support the SRPM is a good effective
model not only for lowest-order level correlations, but also for
correlations on moderate longer ranges. We left an improved study on larger
system size for a future work.

\section*{Acknowledgements}

This work is supported by the National Natural Science Foundation of China
through Grant No.11904069.

\end{document}